\documentclass[reprint,superscriptaddress,amsmath,amssymb,
aps,prb]{revtex4-2}
\usepackage{hyperref}
\hypersetup{colorlinks,allcolors=blue}
\usepackage{dcolumn}
\usepackage{bm}
\usepackage{tabularx}
\newcolumntype{Y}{>{\centering\arraybackslash}X}
\usepackage{wrapfig}
\usepackage{graphicx} 
 \usepackage{bm}

\graphicspath{{pictures/}}

\begin{document}

\title{Phase transitions in rare-erth ferrimagnets with surface anisotropy near the magnetization compensation point}
\author{V.V.~Yurlov}
\affiliation{Moscow Institute of Physics and Technology, Institutskiy per. 9, 141700 Dolgoprudny, Russia} 
\affiliation{New Spintronic Technologies, Russian Quantum Center, Bolshoy Bulvar 30, bld. 1, 121205  Moscow, Russia}
\author{K.A.~Zvezdin}
\email{zvezdin.ka@phystech.edu}
\affiliation{New Spintronic Technologies, Russian Quantum Center, Bolshoy Bulvar 30, bld. 1, 121205  Moscow, Russia}
\affiliation{Prokhorov General Physics Institute of the Russian Academy of Sciences, Vavilova 38, 119991 Moscow, Russia}
\author{A.K.~Zvezdin}
\affiliation{New Spintronic Technologies, Russian Quantum Center, Bolshoy Bulvar 30, bld. 1, 121205  Moscow, Russia}
\affiliation{Prokhorov General Physics Institute of the Russian Academy of Sciences, Vavilova 38, 119991 Moscow, Russia}

\bibliographystyle{apsrev4-2}
\date{\today}

\begin{abstract}
We report of a theoretical model for calculating the H-T phase diagrams of a rare-earth ferrimagnet, taking into account anisotropies originated by  both magnetization sublattices' and by the surface. The possibility of an exchange spring formation due to surface anisotropy is considered. This situation is realized in heterostructures containing a ferrimagnet and a heavy metal. We derive the stability lose lines of the collinear phase from the free energy of the two sublattice ferrimagnet. We numerical calculate the magnetic phase diagrams for the cases when the magnetic field applied along and perpendecular to the easy axis. We demonstrate that tricritical point down at the low field range due to surface anisotropy effect. Moreover, the line of the first order phase transition between angular and collinear phases reduces due to surface anisotropy. In the case when magnetic field is applied perpendicular to the easy axis we show the possibility of the first order phase transition between two collinear phases in contrast to the  phase diagram without surface anisotropy. 
\end{abstract}

\maketitle
\section{Introduction}
Rare-earth-transition metal (RE-TM) compounds is a class of magnetic materials that attracts particular attention in a wide range of different areas such as spintronics\cite{RevModPhys.76.323, doi:10.1146/annurev-conmatphys-070909-104123}, optospintronics\cite{doi:10.1063/1.4976202} and ultrafast magnetism\cite{RevModPhys.82.2731}. This  rare-earth ferrimagnetic (FiM) thin films can be applied for technological uses such as ultrafast memory devices\cite{PhysRevB.95.180409} or high density recording\cite{FERT1999338, doi:10.1063/1.105462, doi:10.1063/1.118202}. Depending on the composition, FiM films may have the magnetization compensation point $T_M$ where the antiferromagneticaly coupled RE and TM magnetizations compensate each other\cite{PhysRevB.96.014409}. This point plays an important role for studying the magnetic phase transitions or magnetization dynamics of the FiM films\cite{doi.org/10.1038/s41578-019-0086-3, PhysRevLett.99.217204, doi:10.1063/5.0010687}.
\par
Studying of the magnetic phase diagrams\cite{ZVEZDIN1995405} is particular of a interest for a better understanding the magnetization dynamics in ferrimagnets. Recent experiments with ferrimagnets such as GdFeCo, GdCo and TbFe  demonstrate anomalous  hysteresis loops near the magnetization compensation point\cite{PhysRevLett.118.117203, OKAMOTO1989259, CHEN1983269, Dy12}. In particular, in the GdFeCo ferrimagnet, triple hysteresis loops are observed above the magnetization compensation temperature \cite{PhysRevLett.118.117203}. At the same time, experiments with TbFeCo with Ta capping layer\cite{PhysRevApplied.13.034053} show that triple loops can appear to the left of the compensation point. To explain this anomalous hysteresis loop, theoretical models\cite{PhysRevB.100.064409} were constructed, in which the interplay of the surface anisotropy, and anisotropies of both sublattices led to a modification of the phase diagrams. The thickness and finite size of a ferrimagnetic film can also significantly affect the spin-reorientation transitions and change the phase diagram. Theoretical \cite{SAYKO1992194, JIANG2014101} and experimental\cite{doi:10.1021/nl301499u} studies on the effect of the surface on the spin dynamics were carried out for example for nanowires\cite{JIANG201395} and for various ferrimagnetic materials\cite{SLONCZEWSKI1992368, doi:10.1063/1.362180}. However, given the new experimental and theoretical results, this area requires further study, and the influence of the surface effects on the FiM phase diagram deserves particular attention. 
\par
In this work we study the magnetic phase diagram for the FiM layer taking into account anisotropies of the both magnetization sublattices and the surface anisotropy. We derive the stability lose lines of the collinear phase from the free energy of the two sublattice ferrimagnet. Influence of the surface exchange anisotropy can be taken into account by introducing the dimensionless parameter which modifies the effective anisotropy of the ferrimagnet. We numerical calculate the magnetic phase diagram for two different directions of the external magnetic field and show the lines of the second and the first order phase transitions. We demonstrate that surface anisotropy can shift down at lower field range the tricritical point and reduce the line of the first order phase transition between angular and collinear phases. In the case when magnetic field is applied perpendicular to the easy axis we show that the surface anisotropy enables the first order phase transition between two collinear phases.

\section{Model and basic equations}
To obtain the magnetic phase diagram for ferrimagnets with the surface anisotropy we use the effective thermodynamic potential\cite{ZVEZDIN1995405}. We assume that transition metal (d-sublattice) is saturated due to large d-d interactions (of the order of $10^6-10^7$ Oe) and rare-earth (f-sublattice) is considered as a paramagnetic in the effective magnetic field. The applicability of this model is substantiated by the hierarchy of exchange interactions\cite{ZVEZDIN1}. Thus, the effective thermodynamic potential without surface anisotropy can be written as:
\begin{equation}\label{eq1}
    \Phi = -\textbf{M}_d\textbf{H} -  \int_{0}^{H_{eff}}M_f(x)dx  - K_a + \Phi_{ex}, 
\end{equation}
where $\textbf{M}_d$ is magnetization of the transition metal sublattice, $\textbf{M}_f$ is magnetization of the rare-earth ions which are saturated in the effective field $\textbf{H}_{eff} = H - \lambda \textbf{M}_d$, $\lambda$ is the f-d exchange constant, $\textbf{H}$ is external magnetic field, $K_a$ is magnetic anisotropy energy and $\Phi_{ex}$ is non-uniform exchange energy. The magnetization of the f-sublattice can be described by the Brillouin function $M_f(x) = \mu_B gJ B_J\Big(\dfrac{g J\mu_B x}{k T}\Big)$, where $g$ is Lande g-factor, $J$ is total angular momentum of the rare-earth ions, $\mu_B$ is Bohr magneton. The anisotropy energy of the ferrimagnet is\cite{PhysRevB.100.064409,PhysRevLett.118.117203}:
\begin{equation}\label{eq2}
    K_a = -K_d\sin^2\psi - K_f\Big(\dfrac{\lambda M_d \sin\psi}{H_{eff}(\psi)}\Big)^2,
\end{equation}
where $K_d$ and $K_f$ are unaxial anisotropy constants of d- and f- sublattices, respectively, $\psi$ is an angle between magnetization of the d-sublattice and the easy magnetization axis. Non-uniform exchange energy can be written as $\Phi_{ex} = A(\nabla \psi)^2$ where $A$ is the exchange stiffness constant. 
Now we should take into account the surface effect as induced exchange magnetic anisotropy. This energy takes the form $\mathcal{F}_{s} = k_d \sin^2\psi_s + k_f(\lambda M_d\sin\psi_s)^2/H_{eff}^2(\psi_s)$, where $\psi_s$ is magnitude of the angle $\psi$ on the surface of the film, $k_d$ and $k_f$ constants of the d- and f- sublattice surface magnetic anisotropy. 
\par
Further expressions is written in the assumption that magnetic film is homogeneous in the plane of the film. We consider the film with thickness $|z| < d$. Exchange surface anisotropy is equal at the edges of the film $k_d(-d) = k_d(d)$ and $k_f(-d) = k_f(d)$ which means that the symmetrical arrangement of magnetization is most beneficial. As a result we can say that $d\psi/dz |_0 = 0$.
\par
Finaly, we obtain the free energy of the ferrimagnet by integrating (\ref{eq1}) over the films volume
\begin{equation}\label{eq3}
\begin{gathered}
    \mathcal{F} = \int_0^d \Phi \;dz + \mathcal{F}_s =  \int_0^d \Big\{A(\nabla \psi)^2 -\textbf{M}_d\textbf{H} -  \\ \int_{0}^{H_{eff}}M_f(x)dx  + K_d\sin^2\psi + K_f\Big(\dfrac{\lambda M_d \sin\psi}{H_{eff}(\psi)}\Big)^2\Big\}dz \\
    + k_d \sin^2\psi_s + k_f\Big(\dfrac{\lambda M_d\sin\psi_s}{H_{eff}(\psi_s)}\Big)^2.
\end{gathered}
\end{equation}
Equation (\ref{eq3}) will be used below to construct the magnetic phase diagram.

\section{Magnetic phase diagram}
We obtain here the magnetic phase diagram of ferrimagnetic material with surface anisotropy using the free energy (\ref{eq3}). We  consider two cases with different directions of the external magnetic field $\textbf{H}$: (a) when the magnetic field is applied along the easy axis and (b) magnetic field is perpendicular to the easy axis. For the case (a) we suppose that $\psi = \theta$ and for the case (b) -- $\psi = \pi/2 - \theta$, where $\theta$ is the angle between magnetization of the d-sublattice and external magnetic film.  Without loss of generality, we consider the case (a) when the magnetic field is aligned with the easy axis. For the case (b), the conclusions given below are also valid. 
\par
We vary the functional (\ref{eq3}) and obtain the Euler-Lagrange equations and boundary conditions to study the free energy:
\begin{widetext}
\begin{equation}\label{eq4}
\begin{gathered}
\Delta\theta = \dfrac{M_d H}{2A}\sin\theta\Big\{1 - \lambda\chi(\theta)  + K_f\Big(\dfrac{\lambda M_d}{H_{eff}(\theta)}\Big)^2\dfrac{1}{M_d H}\Big(2\cos\theta - \dfrac{\lambda M_d H \sin^2\theta}{H^2_{eff}(\theta)}\Big) + \dfrac{2 K_d}{M_d H}\cos\theta \Big\}, \\
\dfrac{d\theta}{dz}\bigg|_s = -\dfrac{k_f}{2A}\Big(\dfrac{\lambda M_d}{H_{eff}(\theta_s)}\Big)^2\sin\theta_s\Big(2\cos\theta_s - \dfrac{\lambda M_d H \sin^2\theta_s}{H^2_{eff}(\theta_s)}\Big) - \dfrac{2 k_d}{2A}\sin\theta_s\cos\theta_s\\
\dfrac{d\theta}{dz}\bigg|_0 = 0,
\end{gathered}
\end{equation}
\end{widetext}
where $\chi = M_f(\theta)/H_{eff}(\theta)$, index $s$ in the second equation defines the film boundary.
\begin{figure}[h!]
\begin{center}
\center{\includegraphics[width=1\linewidth]{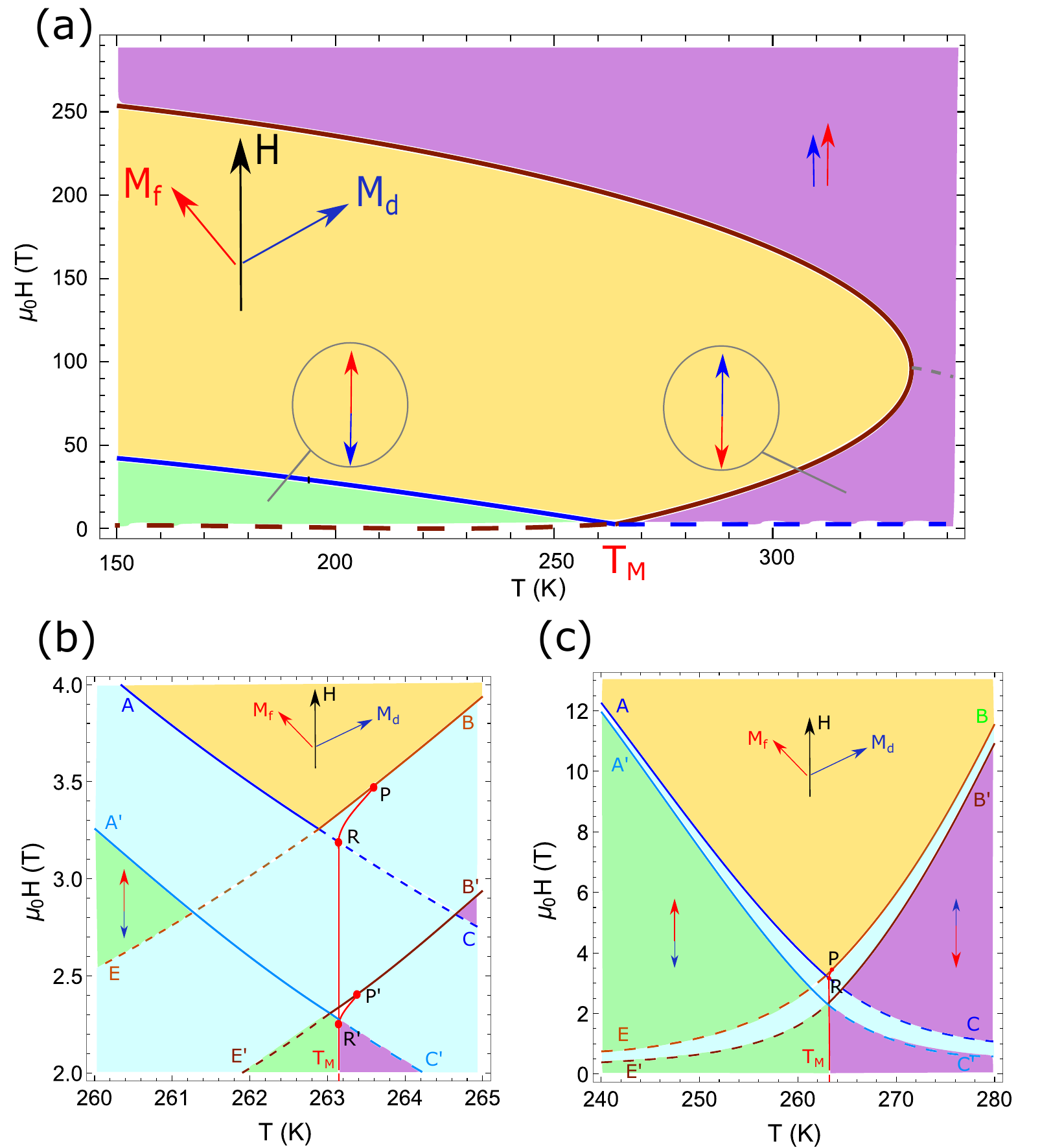}}
\caption{(a) $H-T$ phase diagram of the ferrimagnet in the magnetic field applied along the easy axis in the high field range; solid lines show the second order phase transition between collinear and angular phases, the dotted lines show the second order phase transition between collinear phases; $\theta$ is an angle between external magnetic field and magnetization of the d-sublattice. b) The zoomed-in phase diagram near the tricritical points $P$ and $P'$; line $T_MR'R$ shows the first order phase transition between collinear phases, lines $RP$ and $R'P'$ show the first order phase transition between angular and collinear phases. c) The zoomed-out phase diagram near the magnetization compensation temperature $T_M$. Lines $AC$ and $BE$ is the stability lose lines when the surface anisotropy is zero; lines $A'C'$ and $B'E'$ is the stability in the presence of the surface anisotropy. All diagrams constructed for $0 < h_{eff} < 1$, $K_{eff} > 0$, $k_{eff} < 0$.}
\label{Fig1}
\end{center}
\end{figure}
\begin{figure}[h!]
\begin{center}
\center{\includegraphics[width=0.9\linewidth]{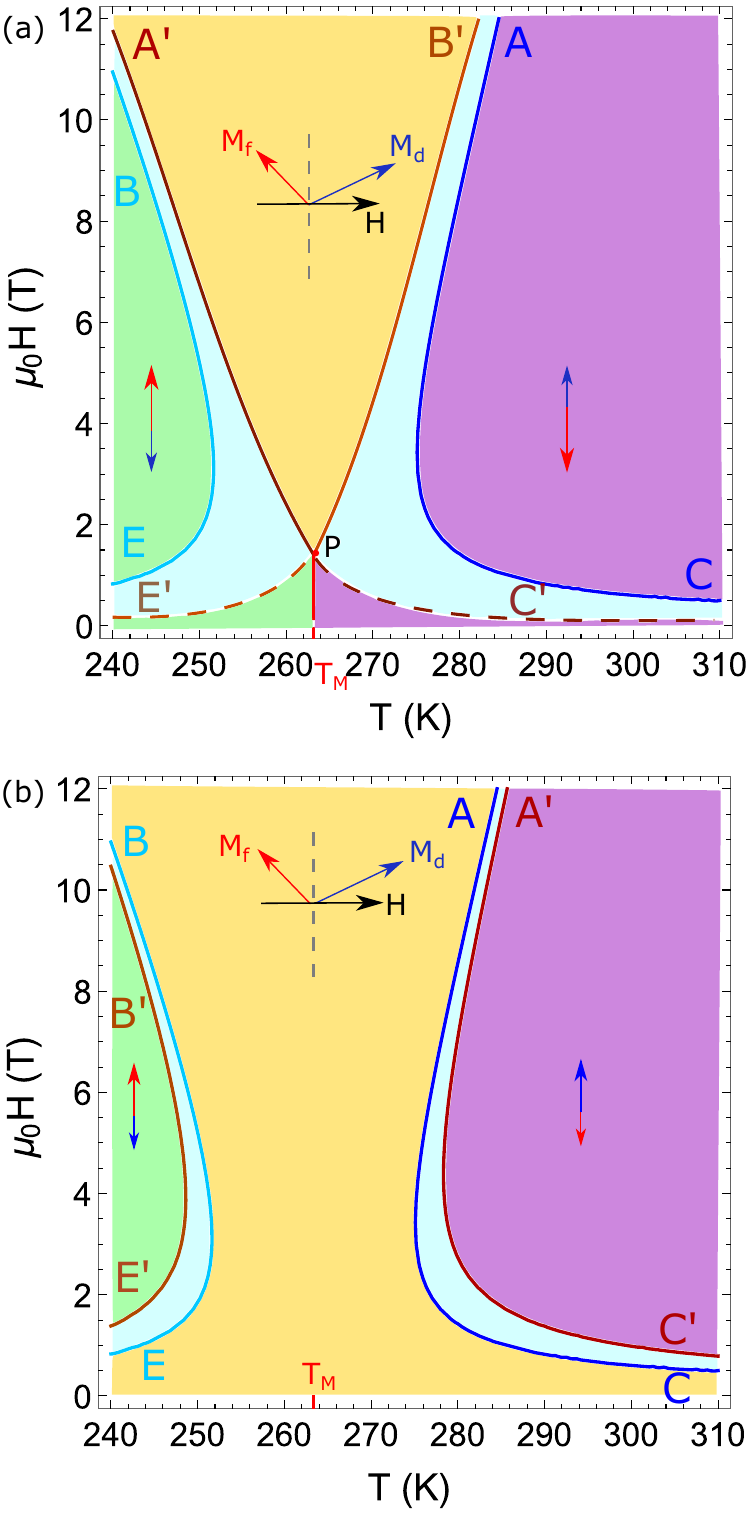}}
\caption{a) $H-T$ phase diagram of the ferrimagnet in the magnetic field directed perpendecular to the easy axis near the magnetization compensation point; solid lines show the second order phase transition between collinear and angular phases, the dotted lines show the second order phase transition between collinear phases; lines $AC$ and $BE$ is the stability lose lines when the surface anisotropy is zero; lines $A'C'$ and $B'E'$ is the stability in the presence of the surface anisotropy; this diagram is constructed for $h_{eff} < -1$, $K_{eff} > 0$, $k_{eff} < 0$; $\theta$ is an angle between external magnetic field and magnetization of the d-sublattice. b) The magnetic phase diagram near the compensation temperature $T_M$; diagram is constructed for $ h_{eff} > 1$, $K_{eff} > 0$, $k_{eff} > 0$.}
\label{Fig2}
\end{center}
\end{figure}
Note that the analytical solution of these equations has some difficulties associated with the definition of the first integral of the differential equation. However, the effect of surface anisotropy can be taken into account by considering the lines of stability loss of the collinear phases. Having this is mind we linearize the (\ref{eq4}) near the lines of stability loss $\theta = 0$ and $\theta = \pi$. Let us give analytical expressions only for the case $\theta = 0$. We look for a solution of the linearized equations in the following form:
\begin{equation}\label{eq5}
    \theta = \overline{\theta}(z)\exp i\{\varkappa_x x + \varkappa_y y\},
\end{equation}
where $\bm{\varkappa}$ is the vector lying in the plane of the magnetic film. We obtain second order differential equation for eigenvalues of the Sturm–Liouville type.
After some calculations we obtain a transcendental expression for the vector $\bm{\varkappa}$:
\begin{equation}\label{eq6}
\begin{gathered}
    \sqrt{\varkappa^2d^2 + \kappa^2d^2}\tanh\{\varkappa^2d^2 + \kappa^2d^2\} = \\ -\dfrac{d}{A}\Big\{k_f\Big(\dfrac{\lambda M_d} {H_{eff}(0)}\Big)^2 + k_d\Big\},
\end{gathered}
\end{equation}
where $\kappa^2 = \frac{M_d H}{2A}\{1 - \lambda\chi(0) + \frac{2 K_f}{M_d H}(\frac{\lambda M_d}{H_{eff}(0)})^2 + \frac{2 K_d}{M_d H}\}$. The stability condition for collinear phases (when magnetizations of both sublattice are parallel and $\theta = 0$ or $\theta = \pi$) is that the equation \label{eq7} has no real solutions. Carrying out similar reasoning for the $\theta = \pi$, we obtain the lines of stability loss of the collinear phases:
\begin{equation}\label{eq7}
\begin{gathered}
    (B'E'): \;1 -\lambda\chi(0) + \dfrac{K_{eff}(0)}{M_d H}(1 - h_{eff}) = 0,\\
    (A'C'): \;1 - \lambda\chi(\pi) -\dfrac{K_{eff}(\pi)}{M_d H} (1 - h_{eff}) = 0,
\end{gathered}
\end{equation}
where $K_{eff} = K_f (\frac{\lambda M_d}{H_{eff}})^2 + K_d$, $h_{eff}$ is the solution of the equation $|\delta_s| = (\frac{\sigma}{h_{eff}})^{1/2}\tanh^{-1}(\frac{\sigma}{h_{eff}})^{(1/2)}$ and $k_{eff} = k_f (\frac{\lambda M_d}{H_eff})^2 + k_d < 0$, $\sigma$ is material surface parameter with positive value, $\delta_s =  (k_f (\frac{\lambda M_d}{H_eff})^2 + k_d)d/A$. 
Similarly, it is possible to obtain lines of stability loss when the magnetic field is perpendicular to the easy axis of the ferrimagnet. For this case, the stability loss lines will be written in the form:
\begin{equation}\label{eq8}
\begin{gathered}
    (B'E'): \;1 -\lambda\chi(0) - \dfrac{K_{eff}(0)}{M_d H}(1 + h_{eff}) = 0,\\
    (A'C'): \;1 - \lambda\chi(\pi) +\dfrac{K_{eff}(\pi)}{M_d H} (1 + h_{eff}) = 0,
\end{gathered}
\end{equation}
where $h_{eff}$ is the solution of the equation $\delta_s = (\frac{\sigma}{h_{eff}})^{1/2}\tanh^{-1}(\frac{\sigma}{h_{eff}})^{(1/2)}$. Here we should note that the present theory applicable in the microscopic range for $d \sim 10^{-7} \div 10^{-6}$ m.  
\par
Recent researches show that anisotropy of the RE sublattice can be larger then the one of the TM sublattice\cite{PhysRevB.100.064409}. Therefore, let us investigate how lines of the first and second phase transitions are changed due to the exchange surface anisotropy. By using the equations (\ref{eq7}), Euler-Lagrange equations (\ref{eq4}), free energy (\ref{eq3}) and methods which are described in \cite{ZVEZDIN1995405} we calculate numerically the magnetic phase diagram of ferrimagnetic film with the exchange surface anisotropy (see in Fig. \ref{Fig1} and Fig. \ref{Fig2}). For the calculations we use the  GdFeCo parameters: $M_d(0) = 4.5 \; \mu_B/f.u.$, $M_f(0) = 7\; \mu_B /f.u.$, $K_d =0.1\cdot 10^5 \; erg/cc$, $K_f = 0.9 \cdot 10^5 \; erg/cc$, $H_{ex} = \lambda M_d \sim 10^6 \; Oe$, $T_M \approx 263 \; K$. Magnetic phase diagrams in Fig. \ref{Fig1} and Fig. \ref{Fig2} represent the areas of the collinear phase where $\theta = 0$ (purple area in Fig. \ref{Fig1} and Fig. \ref{Fig2}), $\theta = \pi$ (green area in Fig. \ref{Fig1} and Fig. \ref{Fig2}) and noncollinear phase $\theta = \theta(T, H)$ (yellow area in Fig. \ref{Fig1} and Fig. \ref{Fig2}). Blue are in Fig. \ref{Fig1} and Fig. \ref{Fig2} shows the different between collinear and noncollinear phases for the ferrimagnet with and without effect of the surface anisotropy. 

\section{Results and discussion}

 Magnetic phase diagram in Fig. \ref{Fig1} and Fig. \ref{Fig2} demonstrate three different phases of the ferrimagnetic film. This phases are: the collinear phase at the high temperature range ($\theta = 0$ purple area), the collinear phase at the low temperature range ($\theta = \pi$ green area) and the noncollinear phase $\theta = \theta(T, H)$ which is indicated as yellow area. 
 \par
Let us discuss now the case (a) when external magnetic field is co-directed with easy axis of ferrimagnet. Fig. \ref{Fig1}(a) show the phase diagram at the high field range. Dotted lines show the second order phase transition between two collinear phases. The zoomed area of the diagram in Fig. \ref{Fig1}(a) is shown in the Fig. \ref{Fig1}(c). Lines $AC$ and $BE$ show the lines of a second-order phase transition between collinear and non-collinear phases. Note that the solid lines denote the second-order phase transition between the collinear and the angular phases. The dotted lines indicate the phase transition between two collinear phases. The grey doted line in Fig. \ref{Fig1}(a) demonstrate the situation when $ H_{eff}(T) = 0$. These lines ($AC$ and $BE$) demonstrate the case when the surface anisotropy is zero ($h_{eff} = 0$). If the surface anisotropy affects the magnetic system, than the effective anisotropy $K_{eff}$ changes, as follows from the (\ref{eq7}) and the $AC$ and $BE$ lines turn into the $A'C'$ and $B'E'$. In Fig. \ref{Fig1}(b) and Fig. \ref{Fig1}(c), the blue color indicates the difference between two cases described above. Fig. \ref{Fig1}(b) and Fig. \ref{Fig1}(c) show that surface anisotropy plays a significant role near low values of the magnetic field $H^* \sim (2K_{eff}\lambda)^{1/2}$. With an increase of the magnitude of the magnetic field, the lines $AC$, $BE$ and $A'C'$, $B'E'$ quickly approach to each other. Fig. \ref{Fig1} is plotted for the $h_{eff} \sim 0.5$. The $T_MR'R$ line in Fig. \ref{Fig1}(b) shows a first-order phase transition line between collinear phases where $\mathcal{F}(0) = \mathcal{F}(\pi)$. The $RP$ line is the first-order phase transition line between the angular and collinear $\theta = 0$ phases. $P$ is the tricritical point. Note that this line is located to the right of the magnetization compensation point $T_M$ due to the influence of the anisotropy of the rare-earth sublattice. Note, that tricritical point $P$ may located to the left from the compensation due to modifying the surface of the ferrimagnetic  by the heavy metal film such as Ta\cite{PhysRevApplied.13.034053}. The line $RP$ transforms into an $R'P'$ due to the exchange surface anisotropy. The first-order phase transition between collinear and noncollinear phases is reduced under the influence of surface anisotropy. Thus, the regions of phase transitions near the compensation point can change significantly due to surface anisotropy. 
\par
The similar situation realises for the case (b)when the magnetic field is perpendicular to the easy axis of the ferrimagnet. Fig. \ref{Fig2}(b) shows that the angular phase expands due to surface anisotropy when the $h_{eff} > 0$. However, the most interesting effect can be seen if the $h_{eff} < 0$. In this case, the spins are pinned on the surface of the ferrimagnet. As a result, the transition between collinear phases in the low-field region can occur through the first order phase transition. This effect is demonstrated in the Fig. \ref{Fig2}(a). In the case described above, the lines $AC$ and $BE$ turns into $A'C'$ and $B'E'$. It also should be noted, that if the anisotropy of the d-subluttice is higher than the one of the f-sublattice than the first order phase transition between the collinear phase $\theta = \pi$ and the angular phase is possible because tricritical point in this case is lower than magnetization compensation temperature. 
\section{Conclusion}
The magnetic phase diagram are studied for the GdFeCo ferrimagnet in presence of the surface magnetic anisotropy for the two different cases: magnetic field applied along and perpendecular to the easy axis. The stability lose lines are derived from the free energy of the ferrimagnet in the assumption that f-sublattice anisotropy is larger than d-sublattice one. In this particular case the tricritical point lies above the compensation temperature. We numerical calculate the phase diagram and show the lines of the second and first order phase transition. We show that in the case when magnetic field is along the easy axis the stability lose lines and tricritical point are falling down in the low field range due to surface anisotropy. Moreover, the area of the first order phase transition between angular and colliniar phase narrows due to surface effects. In other case, when the magnetic field is perpendecular to the easy axis we show the possibility of the realization the first order phase transition between collinear phases due to surface anisotropy. For the both cases which are described above the surface anisotropy change phase diagram significantly only in the low magnetic field. In the high field range the stability lose lines for the cases with surface anisotropy and without one fast approach to each to each other. This findings may be useful for theoretical and experimental study of the spin reorientation transitions.
This research has been supported by RSF grant No. 22-12-00367.
\bibliography{ref}

\end{document}